# Understanding the meaning of Greek temples' orientations.

## Akragas' *Valley of the Temples* as a case study

Robert Hannah

Faculty of Arts & Social Sciences, University of Waikato, New Zealand

Giulio Magli

School of Civil Architecture, Politecnico di Milano, Italy

Andrea Orlando[a,b,c]

[a]Catania Astrophysical Observatory (INAF), Italy

[b]National South Laboratory (INFN), Italy

[c]Istitute of Sicilian Archaeoastronomy, Italy

*The issue of the orientation of Greek Temples has been the subject of several debates since the end of the 19*[th] *century. In fact, although a general tendency to orientation within the arc of the rising sun is undeniable, specific patterns and true meaning remain obscure. With the aim of shedding light on this problem we present here a new complete, high-precision survey of the temples of Akragas, the so-called Valley of the Temples UNESCO site. Our results include a important temple which was essentially yet unpublished, and most of all show that very different reasons influenced the orientation choices – some symbolical, but others by far more practical – besides the general rule of orienting "to the rising sun". In particular, the existence of temples orientated in accordance with the town's grid, as well as to the cardinal points irrespectively from the sun's declination associated to true east at the uneven horizon, is evidenced. Finally, for two temples having "anomalous" orientations a stellar and a lunar proposal are made respectively.*

# 1. Introduction

The ancient Greeks built hundreds of magnificent temples over the course of several centuries, from the seven century BC onward (Lawrence 1996). Leaving aside regional and chronological distinctions in the layout and in the column orders, these sacred buildings were always based on the same conception: an imposing rectangular construction adorned with columns on the facade. Although in many cases the presence of columned porticoes on all sides made the view of the structure enjoyable from all directions, the main principle always remained the same: a Greek temple was meant to occupy a natural place with an obviously man-made feature, and it was to be admired from the outside only.

Admission was reserved to priests and to the privileged few, and public rites were celebrated outside, in front of the temple, which in many cases was equipped with an altar and a platea (religious occasions included festivals, processions and long rituals). The interior of the temple was, strictly speaking, the home of the god. The god's domestic welfare (hence, the beauty and decorum of the building, correct insertion in the landscape, regular giving of daily offerings) was fundamental to assure benevolence and protection to the community. The cult image, located in the central place of the temple, was in many cases an out-and-out masterpiece, like the famous ivory-and-gold statues of Zeus at Olympia and of Athena in the Parthenon in Athens.

The positioning of Greek temples has been the subject of interesting scholarly research. For instance, a connection between the terrain on which the temple is erected and a related deity has been suggested (Retallak 2008). The relationship with the landscape as a whole was first suggested by Vincent Scully (1962). His work pioneered research on the Archaeology of the Landscape, pursuing the idea that landscape and temples formed an architectural unit that was characterized in accordance with the specific god worshiped. In any case, neither the choice of the terrain nor that of the landscape generally implies a specific orientation, so that the matter of the orientation of the Greek temples deserves to be dealt with on its own.

The orientation of a Greek temple is preferably defined as the direction of the main axis from inside looking out, which is the direction in which the statue of the god was in principle looking, as well as being the direction along which the sun would illuminate the facade, which, as we have seen, was the scene for rites and celebrations taking place outside the temple. The majority of these monuments face the eastern horizon, mostly within the arc of the rising sun (Nissen 1873, Penrose 1897, Koldewey and Puchstein 1899, Dinsmoor 1939). However, recent research has shown that eastern orientation is not the universal key to Greek temples, as was previously believed (Boutsikas 2009, Liritzis and Vassiliou 2003, Salt and Boutsikas 2005, Boutsikas and Ruggles 2011, Liritzis and Castro 2013).

What appears to be a simpler situation occurs in the case of the Greek temples of Sicily. The orientation of the temples of Sicily demonstrates indeed a very clear pattern (Aveni and Romano 2000, Salt 2009). It has been determined that 38 out of 41 measured temples are oriented within the arc of the rising sun. This sample is virtually exhaustive for all (but one, studied in this paper) the existing monuments, and clearly we have no need of any statistical analysis to conclude that orientation within the arc of the rising sun was intentional. However, in a way, we are only at the beginning. As it happens, there is no specific concentration of data, for instance, around the solstices or the equinoxes, or other dates for that matter – so how was the alignment chosen? Was it the day of foundation of the temple, or the day of the feast of the god, or what? Perhaps there was a tradition passed down from the original town of provenance? So far, attempts to gain more insight into this problem – for instance, by investigating possible groupings for patron deities – have not been successful. Matters are complicated by the fact that the calendars in use in Greek towns were luni-solar, so that alignments based on feast days would not have been calendrically effective in relation to the timing of the rituals carried out annually (presumably at dawn) in front of the temple. The orientation also appears somewhat unusual when one looks at comparable families of monuments, for instance the Italic temples (temples of the peoples inhabiting continental Italy before the Roman conquest, like the Samnites) which are orientated to the sun ascending in the sky, and the Etruscan temples, which are mostly oriented to the sun ascending or descending in the sky, that is, between the winter solstice sunrise and the winter solstice sunset (Aveni and Romano 1994). It should also be noted that a stellar orientation cannot be distinguished from a solar one if it occurs within the solar arc. Thus all Greek temples oriented to the rising sun also happen to be broadly oriented towards the constellations in which the sun was rising, and can on occasion be accurately oriented to specific stars of such constellations as well as to other stars that had the same declination at the time of construction. A possible, specific interest by the builders in this kind of stellar target must be investigated separately case by case (see eg. Boutsikas and Ruggles 2011).

Motivated by such a variety of open questions, we present here a complete analysis of all the temples of one of the most important ancient Sicilian towns, the world-famous UNESCO site of the Valley of the Temples of Agrigento, ancient Akragas. Our results are actually quite unexpected and show that a variety of factors, not all of them being astronomical, influenced the Akragantine architects.

## 2. The Valley of the Temples

*Akragas* - today's Agrigento - was one of the most important Greek colonies in Sicily, founded in 582 BC by settlers from the nearby Gela and from Rhodes. The site lies on a huge plateau, naturally protected from the north by the Athena Rock and the Girgenti Hills, and from the south by a long rib-hill, bounded on either side by the rivers Akragas and Hypsas, confluent to the south in a single water's course, at the mouth of which the port was constructed.

From the very beginning, under the tyranny of *Phalaris* (570-554 BC), the city was characterized by a regular urban layout, dominated by the Acropolis on the Athena Rock and bordered by the rib Hill which started to house monumental sanctuaries; in the central area were dwellings and public buildings in accordance with a orthogonal grid layout; the necropolis was located outside the city walls. In the last decades of the sixth century BC, Akragas was surrounded by massive walls 12 km long, with 9 gates. The colony reached fame and power under the tyrant *Theron* (488-471 BC), who defeated the Carthaginians at Himera in 480 BC, and during the years of the democracy (471-406 BC) established by the philosopher Empedocles. It is in this period that the extraordinary series of Doric temples, today comprised in the UNESCO archaeological site called *Valley of the Temples*, was built.

There are as many as ten temples in the complex. We list them following the alphabetical order established by Marconi (1939) and using the traditional nomenclature; it is however *fundamental* to recall that only the first two (and the sanctuary of Aesculapius, which is *extra moenia*) are securely attributed to their deities, so that the nomenclature may be misleading.

- Temple of Heracles (A): this is probably the oldest of the Doric temples of Akragas, dating to the end of the 6$^{th}$ century BC. It stands on a mighty pedestal, has a peristyle of 6 x 15 columns (stylobate of 67,04 x 25,28 meters) which is accessed by three steps. The attribution is based on a passage by Cicero who mentions a temple dedicated to the hero and placed not far from the agora, which should likely be in this area.

- Temple of Olympic Jupiter (B): this is the largest Doric temple in the western Mediterranean. The temple was, however, left unfinished and later collapsed, probably due to an earthquake. It is built on a huge stand (56,30 x 113,45 meters) and was reached through a crepidoma of five steps. The most relevant architectural peculiarity comprises the series of figures of stone giants (the *Telamons*) which probably were placed in each intercolumniation. The temple was founded to commemorate the Battle of Himera (480 BC), won by Akragas and Syracuse against the Carthaginians, and it is mentioned by Diodorus and by Polybius.

- Temple of Demeter and Persephone (C): the temple building is now incorporated in the

Norman church of San Biagio. The Doric temple had the cellar preceded by a portico with two columns between the doors. Attribution is based on archaeological material.

- Temple of Juno (D): this temple is located at the eastern end of the hill on a high base. Probably dating to the middle of the 5th century BC, it is hexastyle with a peristyle of 6 x 13 columns, the stylobate measures 38.15 x 16.90 meters while the crepidoma is of four steps. The attribution to Juno (Hera) is purely hypothetical.

- Temple of Concordia (F): this is the best preserved temple in Agrigento. The temple stands on a pedestal with a crawl space inside that corrects the natural inclination of the ground. It is also hexastyle with a peristyle of 6 x 13 columns on a stylobate of 39,42 x 16,92 meters, which is accessed by a crepidoma of four steps. Exceptionally preserved are the architrave with *frieze* of triglyphs and metopes, the *geison*, and, on the fronts, the *tympanum*. Probably almost contemporary to the temple of Juno, with which shares many characteristics, its attribution is completely unknown.

- Temple of Vulcan (G): the temple is peripteral, Doric, also hexastyle, with 13 columns on the long sides. Few remnants are preserved of the building: two stretches of crepidoma with four steps and two columns. The temple is built extending an archaic precinct. Attribution completely unknown.

- Temple of Aesculapius (H): this temple is located outside the city walls, south of the Hill of the Temples, on a substrate of sedimentation clay. The Doric temple has a portico *in antis*, and size of 21.70 x 10.70 meters. It was comprised in a vast sanctuary easily accessible from the sea. A sanctuary to Aesculapius is mentioned by Polybius who placed it outside the walls of the city, so the attribution looks certain.

- Temple of Castor and Pollux (Dioscurides) (I): this is located at the western end of the Hill of the Temples, and thus closes the series. The Doric temple was hexastyle with 13 columns on the long sides (13.40 x 31 meters); today only a reconstructed corner can be seen. The attribution is completely unknown.

- Temple L: adjacent to the temple I, it was left unfinished but the foundations excavated in the rock are clearly visible, together with blocks at the northeast corner, and numerous drums of columns scattered in the building area. Attribution completely unknown.

At least two further temples existed in Akragas. They were located on the Acropolis. One, the so-called Temple of Athena (Marconi temple E) is today buried and only partly visible below the Church of St. Mary of the Greeks. The remains of the second lie almost certainly below the medieval Cathedral nearby.

**3. The orientations of the temples**

The orientations of the Akragas temples were studied for the first time at the end of the 19[th] century. Later, the problem was reconsidered by Aveni & Romano 2000 and Salt 2009. Not all the published data are reliable, however, and not all the temples have been considered (further details below). We have, therefore, re-measured all the temples with a high precision optical theodolite during fieldwork which lasted one week, from 1 to 8 August 2015. Almost all temples have been visited and measured twice in different days. North was calibrated at each measure using a long-distance GPS measure from the theodolite station to a recognizable feature (a corner of a skyscraper) of the modern town of Agrigento (at distances of about 2.5 km); consistency of all measures was cross-checked with compass-clinometer readings corrected for magnetic declination and with Google

Earth readings as well. Due to various reasons, most of all the difficulties in individuating precisely the corners of the side bases of some temples and the fact that some sides are partly covered by huge amounts of sand and/or trees, although the nominal accuracy of the instrument is below 1', we estimate that the error of our measures can be reasonably assumed to be ±15'. Of course we measured also the visible horizon accurately – defined as the visible height from the center of the entrance to the temple – for each temple. It should be noted that in the case of the temples located on the central terrace (temples A, B, I) this was possible only with great difficulty (due to intervening modern features), so that their calculated declinations are only approximate. However, as we shall see, it appears that all these temples were *not* astronomically oriented anyway (Section 4).

As a comparison, it can be noticed that our results *in azimuths* agree relatively close to those of Aveni and Romano who worked with a theodolite, while in some istances they differ from those of Salt obtained with a compass. Our results *in declinations* however differ considerably also from those of Aveni and Romano. We calculated declinations using the program DECCALC by Clive Ruggles, which takes into account atmospheric effects, but this is not enough to explain the difference with the Aveni and Romano's ones. It is unfortunately impossible to further compare the results because these two authors did not give horizon heights; however, a few trials that we have made with their data seem to show that they simply assumed a flat horizon for all temples, a thing which is definitively untrue.

Our data, organized accordingly to increasing azimuths, are the following:

| TEMPLE | AZ | HOR | DEC | |
|---|---|---|---|---|
| TEMPLE L | 77 54 | 2 48 | +11 09 | |
| JUPITER | 78 30 | 2 08 | +10 15 | |
| JUNO | 82 24 | 1 42 | +6 52 | |
| DISCOURIDES | 82 54 | 2 30 | +6 59 | |
| VULCAN | 87 05 | 2 22 | +3 36 | |
| CONCORDIA | 89 36 | 1 08 | +1 33 | |
| AESCULAPIUS | 89 57 | 3 36 | +2 05 | |
| HERACLES | 90 30 | 1 55 | +0 34 | |
| DEMETER | 125 00 | 0 | -26 53 | LUNAR DECL= -27 32 |
| DEMETER(rev) | 305 00 | 2 38 | +28 44 | LUNAR DECL= +29 22 |
| TOWN'S GRID | 78 15 | | | |

To discuss the data, we have divided the temples in groups.

## 4. The temples of the central terrace.

Akragas was planned on the basis on an orthogonal street grid plan in the Greek style, with "meridian" roads (plateai) crossed at right angles by longitudinal streets or stenopoi. We have accurately measured the street grid plan; our result is that the grid is orthogonal with a very good accuracy, and the stenopoi are oriented at 78° 15'. This orientation is probably topographical as it is roughly orthogonal to the slope of the Akragas hill. In particular, a stenopos crossed the hill longitudinally heading towards the central sacred area, which houses the circular sanctuary of the chtonic deities, the temple of the Dioscourides, temple L and, to the left of the road, the temple of Jupiter. There is therefore little (if any) doubt that one of the largest temples of the Greek world, the Akragas temple of Jupiter – azimuth 78° 30' – was orientated *topographically* in accordance with the street grid. Incredible as it may seem, we have been unable to find this simple explanation in the literature.

The same topographical criterion holds for Temple L – azimuth 77° 54' – which fronted the road directly, occupying the horizon of any person descending the hill. This did not occur for the nearby temple of the Dioscourides, which formed instead a sort of scenography for the space fronting Temple L. It is probably for this reason that the Dioscourides temple was skewed clockwise, with an azimuth 82° 54'.

**5. Cardinally oriented temples**

Among the three temples located on the Akragas "rib", the two westernmost ones, Concordia and Heracles, are orientated cardinally, with the front to due east, with a *very good* precision (less than ½° error within our accuracy of 15', so the maximal error committed by the builders certainly did not exceed ¾ °). The same cardinal orientation holds for the sanctuary of Aesculapius, located right below the rib outside of the city walls, which actually furnishes an astonishing 89° 57'. *Not even one of these temples, however, is orientated to the rising sun on equinoctial days*, since a non-trivial horizon raises the declination for each one of them. For Concordia we also verified on site the rising of the sun in alignment with the temple around 25 March; for the other temples the dates are even later and appear scattered without any regularity. We strongly believe, therefore, that these temples were *not* orientated towards the rising sun on specific days of the years. We propose here a new idea, namely that they were oriented *cardinally*. Astronomy was there, of course, in determination of true east (or, more likely, of true north) and symbolism was certainly there as well, with the choice of orienting a square sacred building with the sides along the cardinal directions. However, the builders – who were of course aware that the sun rises at true east on the equinoxes only if the horizon is flat – were not interested in this last phenomenon, which might have occurred causally if their horizon happened to be flat.

In addition, the Temple of Vulcan (azimuth 87° 05') may probably be added to the cardinal family, in spite of its huge deviation from true east. In fact the temple sits on a separate hill with the narrow

Colimbetra valley passing behind. The temple contains - and is built over - an archaic building oriented along the line of maximal slope of the hill ~80°, and the project of the enlarged monument was probably skewed as much as possible towards the cardinal directions.

**6. The Temple of Juno**

This magnificent temple sits at the easternmost end of the rib, and is sustained by a huge artificial terrace. To begin the study of this temple, we verified accurately and without any possible doubt the following fact: placing the very same building with a cardinal orientation was possible without any geological or topographical obstruction and over the very same terrace. Therefore, there must be a different reason why this temple is not orientated cardinally, as the two which follow along the rib instead are. We propose here the possibility that this temple was orientated to the stars. The azimuth of the temple 82° 24', in the fifth century BC, indicates very neatly the region of the sky where a relatively faint constellation (its brightest star is of magnitude 3.8), Delphinus (the Dolphin) was rising. As is obvious, it is impossible to speak about the azimuth of rising of an entire constellation, so this assertion needs a detailed explanation. Delphinus occupies a small portion of the sky, which can be individuated by a small "quadrilateral" of four stars. Due to their high magnitude, we must of course consider these stars at their minimal visible height due to extinction, which according to Thom's rule (visible height = magnitude in degrees) is not less than about 4°. In the Akragas sky of, say, 450 BC these four stars were visible in the region of azimuths between 80° 45' and 83° 15', with the unique other relevant star of the constellation, Epsilon, at 86°. We have therefore a very good match with the Juno temple in the century of its construction.

The attribution of the temple is completely unknown, and an orientation to Delphinus – in spite of the faintness of the constellation – could make sense, if a dedication to Apollo (a god who is absent from any other temple in Akragas) could be suggested. Delphinus is indeed one of the constellations connected to Apollo mythology and, according to a recent proposal, was used as a marker for the season of the pilgrimage to Delphi (Salt and Boutsikas 2005).

## 7. The Temple of Demeter and Persephone (San Biagio)

The temple (labelled C) is of Doric style, and belongs to the final phase of the Archaic period (480-470 BC). It is very well preserved because in the Middle Ages the building was transformed into a church dedicated to San Biagio. The facade of the church points to the north-west, but it is very likely that the facade was obtained by opening an entrance in the back wall of the cellar of the temple, which was therefore originally fronting south-east. However, of the portico with two columns that the Greek temple should have had, only the (so-called honeycomb ) foundation sectors can be seen. The stone foundation platform of the whole building and a part of the original stone walls is preserved, accurately built in isodomic masonry of huge rectangular blocks. Inside the church, excavations have revealed a cistern belonging to the Greek phase, located close to the north-west corner and therefore inside the cellar of the temple, a quite unusual feature (Bellavia et al. 2012).

There is no available space for an altar in front of the temple, since the terrain slopes down abruptly on to the escarpment of the city walls. A relatively large esplanade is instead present on the back (that is, in front of the church). This esplanade is contemporary to the temple and was obtained artificially through the construction of huge retaining walls on the south side and an accurate excavation and leveling of the rock on the north side. The area was accessed from the town through a large road partly excavated in the rock as well, which is still today perfectly visible.

During the excavations of 1925 a votive deposit was found, formed by a large amount of objects. In particular, there were many fragments of two female busts of terracotta, one of which could be reconstructed in its integrity and was identified as Persephone. As a result of this discovery the temple is attributed to the Eleusinian divinities. The attribution to Demeter and Persephone was also confirmed by the presence of two small circular altars: one is solid , with a diameter of 2.53 meters, and the other - with a diameter of 2.70 meters - has a central well (*bothros*), which was found filled with ritual offerings, i.e. broken *kernoi*, or ritual vessels of Demeter. These altars are located in the

"corridor" formed between the rock cut to the north and the side of the temple.

The front of the temple could be accessed directly by a stairway which crosses the town's walls through a postern and leads to a strange building located *extra moenia.* It is a protohellenic (7[th] century BC) sanctuary probably dedicated to chthonic deities, whose architectural elements are integrated with the natural features of the site, as is often found in the holy shrines of the gods of the earth (for example at Eleusis, in Lykosoura or in Enna). The sanctuary consists of a rectangular building up on the cliff (below the temple of Demeter and Persephone), on which there are two communicating hypogea, which were filled with votive offerings. A third gallery was used as an aqueduct to supply water collected from a nearby source in the basin of the building, which was therefore a sort of fountain-sanctuary (Zoppi, 2005). Recently it has been proposed to connect the rock sanctuary to nymphal cults (Portale, 2012) and new studies are underway to better understand its enigmatic architecture (Fino, 2014).

The temple of Demeter and Persephone was measured for the first time by Heinrich Nissen in his book *Das Templum* (Berlin, 1869) and by Koldewey and Puchstein (1899); for reasons we do not know, it has passed unnoticed in more recent works. Nissen was a very serious scholar and his data are usually reliable, but not in this case. In fact, to our surprise, we discovered that the azimuth given by Nissen, which is the same as the rising sun at the winter solstice, is in defect by as much as 4 degrees. As a consequence, the true azimuth of the temple *falls below the arc of the rising sun*. To our surprise, then, we discovered that the temple is the only one in Akragas whose declination is not in the solar range[1], and adds to the very few Sicilian temples whose declination has this property. The horizon is flat, and the declination is very close to -27°.

It can be noticed that the building could have been skewed some 4 degrees towards the east on its platform, in this case in order to align with winter solstice sunrise, without any practical problem. Even more, although the horizon in front of the temple is flat, a rock curtain located immediately to the left (east) looking from the entrance was left in situ (the huge excavation of the terrace ends just nearby), and this curtain even obscures the midwinter sun at rising.

---

[1] It is possible to estimate the azimuth of the basement of the Athena temple on the Acropolis at ~110°.

We conclude then that the temple is not aligned with the rising sun (of course, the back is also not aligned with the setting sun). Perhaps this fact seemed so unnatural to Nissen that he decided to be wrong in his measures and adjusted it to the closer solar azimuth? We shall probably never know. In any case, the – clearly deliberate – orientation of the temple cries out for an explanation.

The first thing to be noticed is that declination -27° is enclosed between the declinations of the sun at winter solstice and that of the moon at the major southern standstill, and fits the Venus minimal declination very precisely. We were therefore at first intrigued by the idea that the temple could be aligned to Venus. However, although Venus can attain – in principle – its maximal and minimal declinations both as the morning and as the evening star, the morning star has *never* had a declination significantly greater than that of the sun at the solstices in the last four millennia or so (Sprajc 1993). This fact is already well known from studies on Venus alignments in the Mayan world, but in any case we verified it explicitly and independently in all the 8-year Venus cycles of the 5[th] century BC. No conspicuous stars or asterisms correspond to this declination in this period either.

We then re-analysed the orientation taking into account the possibility of a lunar alignment. Lunar declinations are affected by parallax by some ½ ° and therefore the front of the temple – again using the program DECCALC kindly provided by Clive Ruggles – yields a lunar declination -27° 32', not a significant value. At this point, we visited the temple again in search of an explanation, and we actually found one which we propose here, although we are well aware that it can be only the result of a progressive selection of the data made by ourselves.

It is very likely that processional rites were carried out which involved both the fountain sanctuary and the temple uphill, and is very *unlikely* that any procession traveled *downhill* from the temple to the fountain. So, we can imagine a nocturnal procession coming up from the sanctuary and reaching the temple, in front of which, however, there is not – and there never has been – enough space to house worshipers. It is therefore conceivable that the people, after the ascent, crossed the corridor between the north side of the temple and the hill (perhaps throwing votive offerings in the bothros) and gathered in the vast esplanade located on the *back* of the temple (recall that this esplanade has been constructed artificially and with huge effort). In our new visit we accurately measured the horizon to the north-west from the facade of the medieval church and therefore from the back side of the temple. This horizon is very striking, since it is occupied by the hill where the acropolis of Akragas once sat. The tower of the medieval cathedral, which should correspond to the main temple on the acropolis, is clearly visible directly in front of the temple. Measuring the horizon we took into account an estimated average for the height of the modern buildings, and the results in declination are +28° 44' uncorrected by parallax, with a lunar declination +29° 22'. This result is impressively close to the maximal lunar declination which in the 5$^{th}$ century BC (due to the slight variation of the obliquity) was around 28° 50'.

As is well known, precise lunar extreme azimuths are very difficult to individuate, but the full moon near to the winter solstice in the years close to the standstill attains an azimuth which is always very close to the maximal one. All in all then, we propose a lunar connection for the complex of San Biagio, a fact which would fit well with the religious connections of the water cult in the Mediterranean basin. A connection might exists also with the choice of devoting the Christian church to San Biagio, a saint who blessed waters and animals (Barra Bagnasco, 1999).

## 8. Discussion and conclusions

We believe that our results for Akragas, besides having interest *per se*, show that the orientation of the Greek temples of Sicily could be affected by different considerations, so that there is no "golden rule" to explain it. In particular, the long sought "day=god" rule appears to be rather a superimposed idea which is probably due to a misleading parallel with (some) Christian churches. At least in the case of the temples of Akragas, reasons coming from urban layout and/or morphological aspects of the terrain could be as important as symbolical ones. On the other hand, symbolical reasons could be more difficult than expected: in particular, our data clearly show the existence of Greek temples orientated along the cardinal directions *irrespective* of the solar date to which they would match due to the horizon. For such temples, only a general rule imposing the facade "towards the eastern horizon" was applied. Finally, although more speculative, a stellar orientation and a lunar one appear as the most likely explanations for the two temples (one of which is essentially unpublished) where the construction of an artificial terrace would have easily allowed a "more standard" choice.

**Acknowledgements**

The present work has been developed within the convention for scientific cooperation existing between the "Ente Parco Archeologico della Valle dei Templi" and the FDS Laboratory for the Communication of Science at the Politecnico of Milan. The authors wish to thank the director of Ente Parco, Arch. Giuseppe Parrello, and the responsible Dr. Giuseppe Presti for their continual cooperation and support.

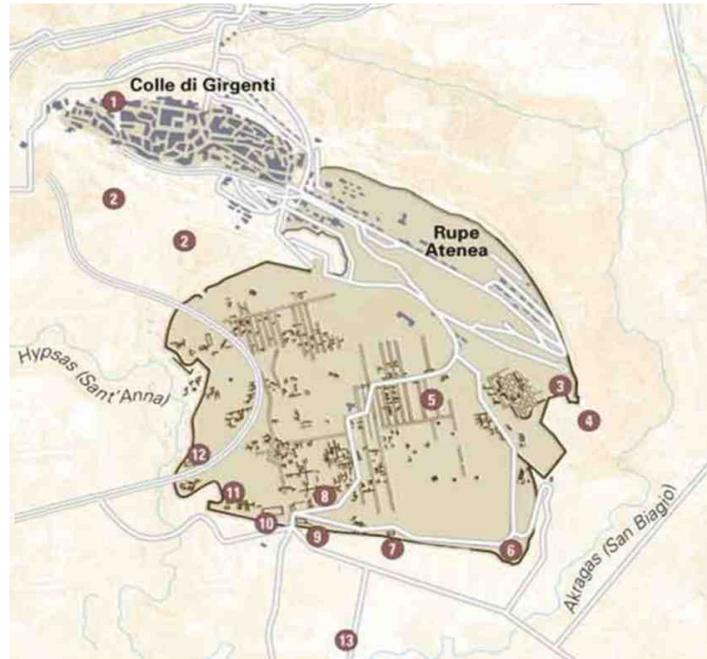

Fig.1. Plan of ancient Akragas (town's wall in black). 1) Athena Temple 2) Necropolis 3) Demetra Temple 4) Archaic Sanctuary 5) Town blocks 6) Juno Temple 7) Concordia Temple 8) Agora'  9) Temple of Heracles 10) Temple of Jupiter 11) Temple of Disocurides and Temple L 12) Temple of Vulcano 13) Temple of  Aesculapius

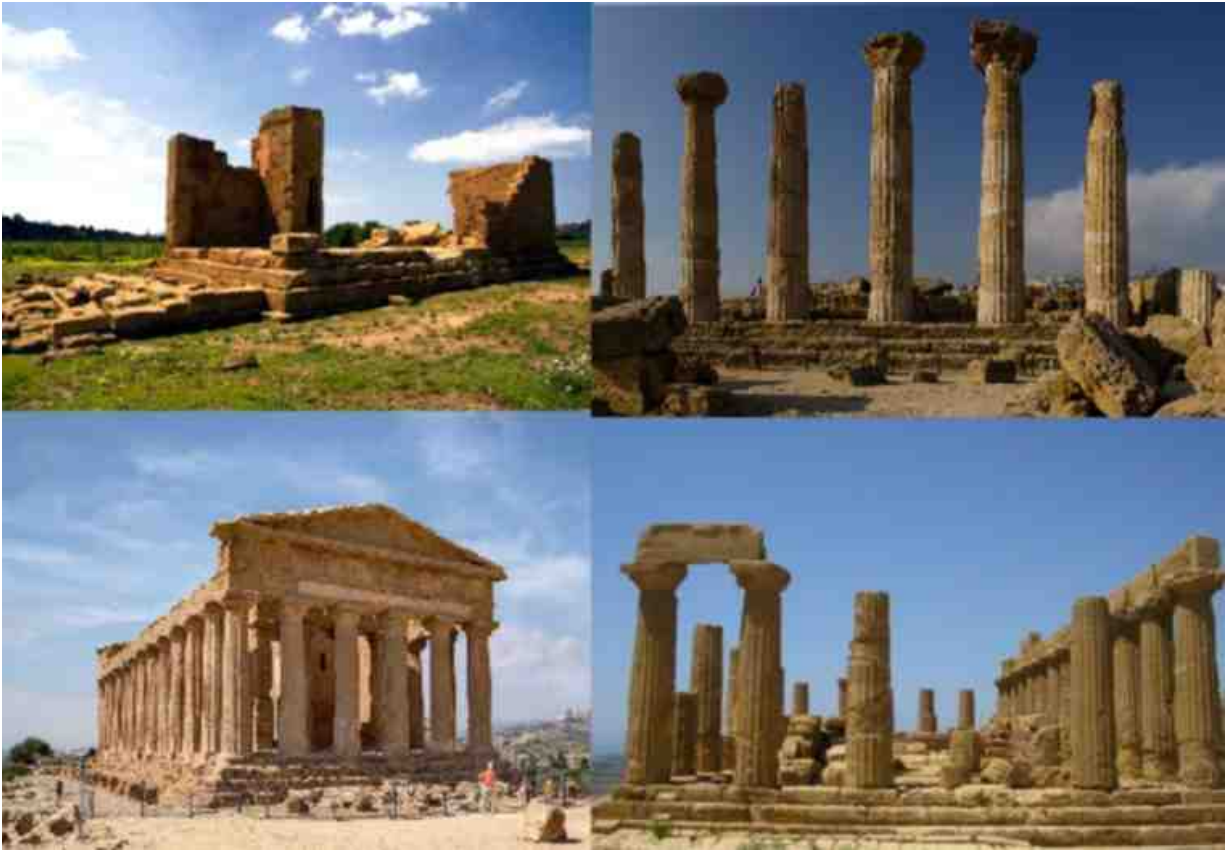

Fig. 2. Akragas, the Temples. Anti-clockwise from upper left: Aesculapius, Heracles, Juno, Concordia.

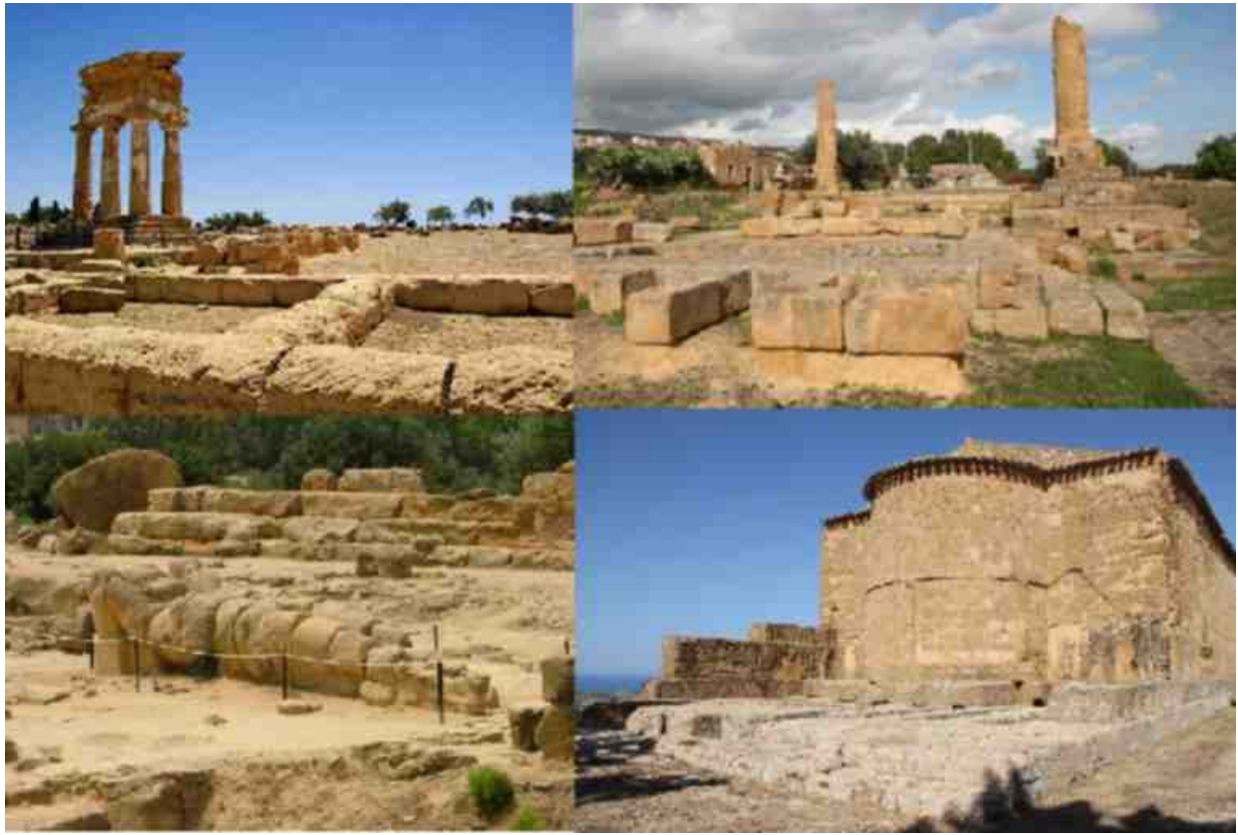

Fig. 3. Akragas, the Temples. Anti-clockwise from upper left: Dioscurides and Temple L foundations, Vulcan, Demetra, Jupiter.

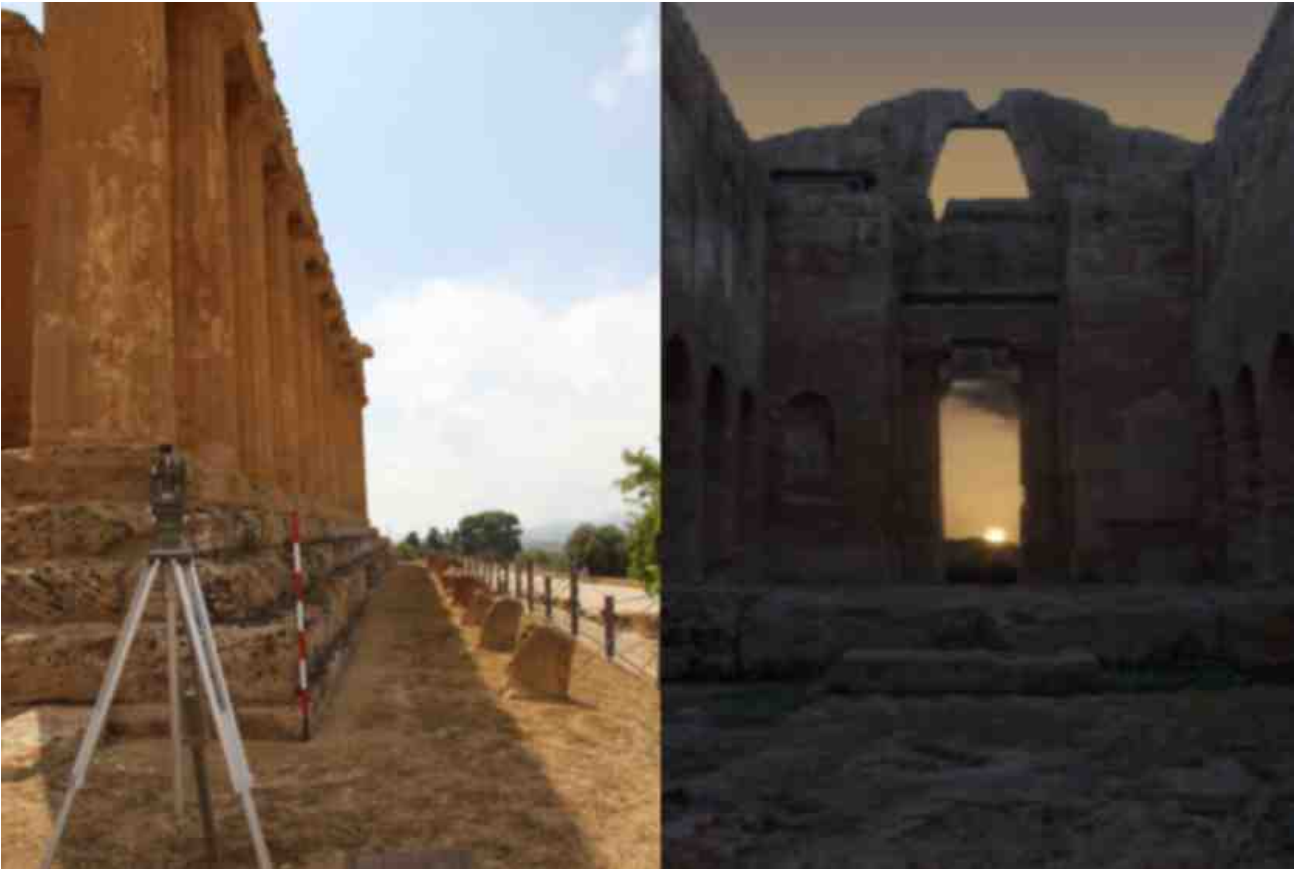

Fig. 4. Akragas, Concordia Temple. Left: measuring the north side; right: the sun rises in alignment with the axis of the temple on March 25, 2015.